\documentclass[reprint,NumberedRefs]{JASA} % JASA.cls
\usepackage{amsmath,amssymb,mathtools,bm}
\usepackage{graphicx}
\usepackage{url}
\usepackage{booktabs}
\usepackage{enumitem}
\usepackage{tabularx}
\usepackage{array}
\usepackage{siunitx}

\newcommand{\bs}{\sffamily}

\begin{document}

\title[Diffuseness estimation with tight-frame arrays]
%{A Unified Framework for Direction and Diffuseness Estimation Using Tight-Frame Microphone Arrays}
{A framework for diffuseness evaluation using a tight-frame microphone array configuration}

\author{Akira Omoto}
\affiliation{Faculty of Design, Kyushu University}
\email{omoto@design.kyushu-u.ac.jp}

\date{\today}

\begin{abstract}
This study extends the theoretical framework of diffuseness estimation to practical microphone arrays with different spatial configurations. Building upon the covariance-based model, we formulate a velocity-only covariance framework that enables consistent diffuseness evaluation across heterogeneous array geometries without requiring mode whitening or spherical harmonic decomposition. Three types of arrays --- the A-format, rigid-sphere, and a newly proposed tight-frame array --- are modeled and compared through both simulation and measurement experiments. The results demonstrate that the proposed tight-frame configuration achieves isotropic sampling and reproduces diffuseness characteristics equivalent to those of higher-order spherical arrays, while maintaining a compact physical structure. 
We also examine the accuracy of acoustic-intensity detection within the same framework. 
The findings bridge theoretical diffuseness analysis and practical implementations, contributing to the design of robust sound-field measurement systems.
\end{abstract}

\maketitle

\section{Introduction}

A diffuse sound field is an ideal state in which both sound pressure and particle velocity are spatially and directionally isotropic.  
This assumption underpins the classical reverberation theories and remains central in architectural acoustics. 
Quantitative evaluation of “diffuseness” thus plays a key role—not only for objectively assessing sound fields but also in applications such as room design, spatial reproduction, and auditory analysis.

Historically, early studies derived diffuseness indicators from spatial fluctuations of reverberation time~\cite{Schroeder1959} and from statistical limits in frequency-domain descriptions of sound fields~\cite{Schroeder1962}.  
Theoretical advances introduced diffusion factors to describe energy homogeneity~\cite{Schultz1971} and correlation-based coefficients~\cite{Bodlund1976}.  
The spatial correlation of sound pressure in a diffuse field was shown to follow a sinc-type function~\cite{Koyasu1971}.

In parallel, Thiele visualized reflection directions and strengths in concert halls, linking diffuseness with perceptual attributes such as clarity~\cite{Thiele1953,Meyer1954,Meyer1956}.  
Following this, Gover and colleagues~\cite{Gover2002,Gover2004} developed spherical microphone-array systems capable of analyzing the directional and spatial variations of reverberant fields. 
%By steering beams over multiple directions, they computed the variance of directional power to define “directional diffusion” and the anisotropy index, showing how diffuseness evolves over time as the reverberant field becomes more isotropic and later anisotropic.  
%These studies established the idea that a diffuse field can be characterized by the \emph{uniformity of directional energy distribution}, forming an experimental foundation for later diffuseness metrics.
By steering beams over multiple directions, they introduced anisotropy and directional diffusion measures based on the variance of directional power, establishing the idea that diffuseness can be characterized by the uniformity of directional energy distribution.

In the 2000s, diffuseness quantification in time and space advanced rapidly.
The First Order Ambisonics (FOA) based intensity-energy ratio (I/E) method by Merimaa and Pulkki~\cite{Merimaa2005,Pulkki2006} characterized sound fields through the relationship between acoustic intensity and energy density.
Acoustic intensity was obtained from impulse responses, allowing the estimation of diffuseness and major reflection directions, later reproduced using Vector Base Amplitude Panning (VBAP)\cite{Pulkki1997} in Spatial Impulse Response Rendering (SIRR).
The same principle was extended to continuous signals in Directional Audio Coding (DirAC)\cite{Pulkki2007,Vilkamo2009}.

In parallel, diffuseness measures based directly on the statistical properties of acoustic intensity vectors were proposed.
In particular, Ahonen and later Del Galdo et al. introduced indicators derived from the coefficient of variation of the intensity magnitude, relating diffuseness to the ratio between the norm of the mean intensity vector and the expected instantaneous intensity~\cite{Ahonen2009,DelGaldo2012}.
%These CV-based measures rely solely on first-order acoustic quantities and offer a complementary perspective to I/E-based formulations.

Diffuseness estimation then became closely linked with spherical microphone arrays and spherical-harmonic analysis~\cite{Rafaely2015}.  
Epain and Jin~\cite{Epain2016} made a significant contribution with the covariance matrix eigenvalue diffuseness estimation, abbreviated as COMEDIE, method, which interprets diffuseness as a function of the eigenvalue distribution of spherical-harmonic signal covariance.  
In addition, they formulated a set of benchmark scenarios—such as single plane-wave, mixed diffuse-plus-direct, and interference fields—that have since served as reference frameworks for validating diffuseness estimators.  

Following the covariance-based formulation by Epain and Jin\cite{Epain2016},
Politis and Pulkki\cite{Politis2015,Politis2016} extended this concept to spatially weighted energy analysis and parametric spatial audio frameworks.
Their work unified the estimation of intensity, energy density, and diffuseness in the HOA domain.
More recently, McCormack et al.~\cite{McCormack2020} investigated the perceptual effects of spherical order, dedicated diffuse rendering, and frequency resolution in higher-order Spatial Impulse Response Rendering (SIRR), highlighting that perceived spaciousness strongly depends on the accuracy of diffuse-field modeling.

Throughout this history, the link between diffuseness estimation and intensity measurement has been widely recognized, as clearly demonstrated by the I/E method.  

In addition to the conventional pressure-pressure method, the principle of acoustic intensity measurement using oppositely oriented cardioid microphones was first proposed by Bauer~\cite{Bauer1968}, forming the basis of what is now known as the C-C method.
This approach was later refined and systematically established by Hoshi and Hanyu~\cite{Hoshi2018}, also Hanyu and Hoshi~\cite{Hanyu2024}, who developed reliable formulations for practical intensity measurements in various acoustic environments.
These studies advanced an integrated view of intensity and diffuseness, linking energetic analysis with spatial measurement.

In the present study, we aim to bridge theoretical diffuseness analysis and practical microphone measurements.  
A multichannel microphone array is employed to estimate broadband acoustic intensity simultaneously.  
By adopting uniform directional sampling based on a tight-frame array, the proposed method achieves both a higher usable frequency and improved stability.  
Combined with impulse-response measurements, it enables high-resolution evaluation of spatial energy distributions, including reflections.

\vspace{5mm}

Building upon these theoretical foundations, the present study extends diffuseness estimation to practical microphone arrays with the following specific contributions:

\noindent
\textbf{Microphone array based on tight-frame theory and spatial sampling} ---
A tight frame provides a redundant yet stable sampling strategy on the sphere, allowing uniform directional coverage without requiring exact spherical harmonic reconstruction. We propose a new microphone array shape based on this theory.

\noindent
\textbf{Velocity-only covariance framework} ---
A formulation that allows consistent diffuseness evaluation without relying on mode decomposition or ideal spherical harmonics, thus enabling unified analysis across heterogeneous array geometries.

\noindent
\textbf{Systematic comparison across indices and array configurations} ---
Three array types: A-format, Tight Frame, and Rigid Sphere --- are compared under identical conditions to evaluate consistency, frequency dependence, and the effects of spherical-harmonic order and whitening. 
Emphasis is placed on the trade-off between the numerical instability of high-order spherical harmonic (HOA) analysis \cite{Epain2016} and the implementation stability of first-order (FOA) approaches.
Overall, these analyses aim to bridge theoretical validity and practical reproducibility, thereby establishing a consistent, practically applicable framework for diffuseness evaluation.

%%%%%%%%%%%%%%%%%%%%%%%%%%%%%%%%%%%%%%%%%

\begin{figure}[b]
  \centering
  \includegraphics[width=0.7\linewidth,pagebox=artbox]{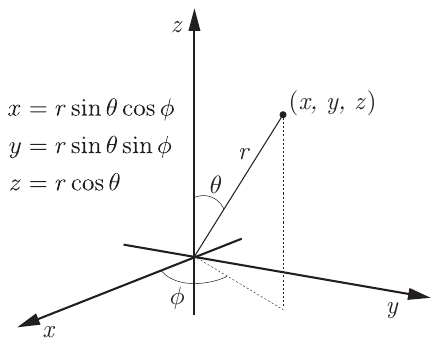}
  \caption{Coordinate system used in this study.}
  \label{fig:RHS}
\end{figure}

\section{Microphone arrays for diffuseness assessment}

Diffuseness estimation mainly relies on acoustic intensity and energy density, as well as on eigenvalue analyses of covariance matrices of spherical-harmonic (SH) components.
Since both contain spatial and directional information, accuracy and effective bandwidth depend strongly on the geometry of the microphone array.  
Commonly used devices include p-p probes, p-u sensors, paired cardioid microphones, and spherical arrays such as the Eigenmike\textsuperscript{a)}.
In particular, A-format microphones and rigid-sphere arrays are widely used for practical analysis.

An A-format microphone often conceals the directional responses of its capsule, hindering quantitative validation.  
Conversely, rigid-sphere arrays enable high-order SH expansion but require many channels, complex calibration, and suffer from numerical instability at high orders.  
Hence, there is a growing need for arrays that achieve uniform spatial sampling while remaining simple and robust.

In this study, we model both the A-format and rigid-sphere microphone arrays and further propose a stably configured array referred to as a tight-frame arrangement, examining the characteristics of each. 
The coordinate system used in the following discussion is shown in Fig. \ref{fig:RHS}.

\begin{figure}[t]
  \centering
  \includegraphics[width=0.54\linewidth,pagebox = artbox]{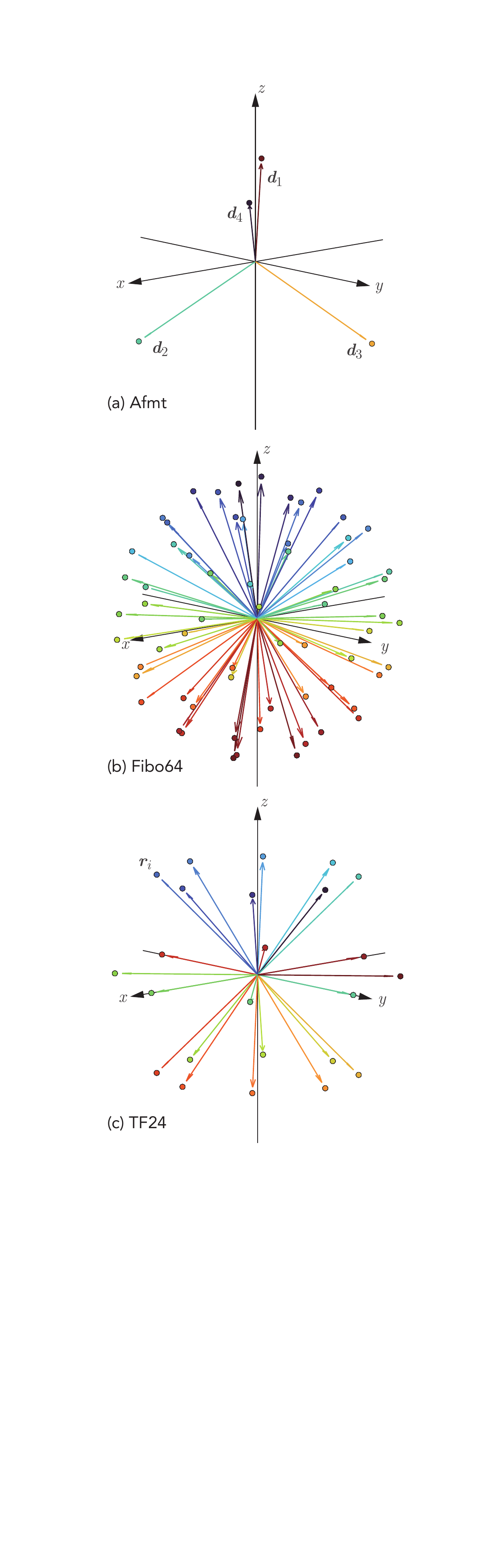}
  \caption{Microphone arrangements used in this study. (a) Microphone arrangement modeled after the AMBEO VR Mic (A-format array).
(b) Array model based on the Eigenmike, in which 64 omnidirectional microphones are uniformly distributed on a rigid-sphere baffle according to a Fibonacci lattice.
(c) Proposed tight-frame microphone configuration, in which microphones are spaced every 45° in azimuth and ±45° in elevation.}
  \label{fig:mic_directions}
\end{figure}

\subsection{A-format microphone (Afmt)}

An A-format microphone uses four capsules in a regular tetrahedral configuration, from which the FOA components are derived for straightforward spatial analysis.  
This configuration, referred to as {\bs Afmt}, inherently offers limited spatial resolution and reduced directional discrimination at high frequencies.

In the simulations, the A-format microphone is modeled after the AMBEO VR Mic \textsuperscript{b)}, consisting of four cardioid capsules oriented along the axes shown in Fig.~\ref{fig:mic_directions}(a) within the coordinate system of Fig.~\ref{fig:RHS}:
\begin{alignat*}{2}
\bm{d}_1 &= \frac{1}{\sqrt{3}}(+1,+1,+1) &\;\; &: \;\; \text{FLU (Front Left Up)}, \\
\bm{d}_2 &= \frac{1}{\sqrt{3}}(+1,-1,-1) & &: \;\; \text{FRD (Front Right Down)}, \\
\bm{d}_3 &= \frac{1}{\sqrt{3}}(-1,+1,-1) & &: \;\; \text{BLD (Back Left Down)}, \\
\bm{d}_4 &= \frac{1}{\sqrt{3}}(-1,-1,+1) & &: \;\; \text{BRU (Back Right Up)}.
\end{alignat*}

Let $D_1$ to $D_4$ be the complex amplitude of measured signals.  
They are converted into the B-format signals $(W,X,Y,Z)$ as
\begin{equation}
\begin{bmatrix}
W \\ X \\ Y \\ Z
\end{bmatrix}
=\frac{1}{2}
\begin{bmatrix}
+1 & +1 & +1 & +1\\
+1 & +1 & -1 & -1\\
+1 & -1 & +1 & -1\\
+1 & -1 & -1 & +1
\end{bmatrix}
\begin{bmatrix}
D_1 \\ D_2 \\ D_3 \\ D_4
\end{bmatrix}.
\label{eqn:AtoB}
\end{equation}

The corresponding sound pressure and particle velocity are
\begin{equation}
p=W,\quad
\bm{u} = -\frac{\sqrt{3}}{Z_0}
\begin{bmatrix}
X \\
Y \\
Z
\end{bmatrix}
\label{eqn:pu}
\end{equation}
where $Z_0$ is the characteristic impedance of the medium.  
From these quantities, both intensity and the total energy can be derived as
\begin{equation}
\bm{I}=\frac{1}{2}\Re\{\,p\,\bm{u}^\ast\,\},
\quad
E=\frac{|p|^2}{4\rho_0 c^2}+\frac{\rho_0}{4}|\bm{u}|^2,
\label{eqn:IandE}
\end{equation}
where \(\Re\{\cdot\}\) denotes the real part and superscript \((\cdot)^{*}\) denotes complex conjugation. Also, $\rho_0$ and $c$ represent the mass density of the medium and the sound speed, respectively.
The ratio $|\bm{I}|/(cE)$ serves as the diffuseness index in the I/E method.

\subsection{Rigid-sphere baffle microphone array (Fibo64)}

Spherical arrays such as the Eigenmike place microphones on a rigid sphere and use SH analysis to decompose the sound field into higher-order components, enabling analysis of both overall diffuseness and local anisotropy.  
Practical challenges remain:
\begin{itemize}
\item Accurate modeling of the sphere and baffle acoustics is required.
\item Numerical stability degrades at high orders.
\item Dense sampling increases calibration and cost.
\end{itemize}

Thus, despite their theoretical power, spherical arrays face trade-offs between spatial uniformity and implementation simplicity.

In this study, a 64-channel array of omnidirectional microphones uniformly distributed on a rigid sphere (radius 0.042 m) is simulated.  
From the simulated pressures, sound pressure, particle velocity, and HOA coefficients are computed following Rafaely\cite{Rafaely2015}.

\medskip\noindent
\textbf{Sound pressure on a rigid sphere}  

The pressure on a rigid sphere of radius $r$ can be expanded as
\begin{equation}
p(\Omega,k)
=\sum_{n,m} a_{nm}(k)\, b_n(kr)\, Y_{nm}(\Omega),
\label{eq:rigid_response_simplified}
\end{equation}
where $b_n(kr)$ is the radial filter corresponding to the rigid-boundary condition for the wavenumber $k$.  
Each microphone measures the pressure at direction $\Omega_q$.

Let $\bm{p}=[p_1,p_2,\dots,p_Q]^{\!\top}$ denote the pressures measured at $Q$ positions $\{\Omega_q\}$.  
The discrete model is then expressed as
\begin{equation}
\bm{p} = \bm{Y}\, \bm{B}\, \bm{a},
\label{eq:discrete_model2}
\end{equation}
where $\bm{Y}$ is the spherical harmonic (SH) matrix, $\bm{B}$ is the diagonal matrix of radial filters, and $\bm{a}$ is the vector of higher-order coefficients.

\medskip\noindent
\textbf{Estimation of HOA coefficients}

The estimation of HOA coefficients $\widehat{\bm{a}}$ are then estimated in the least-squares sense as
\begin{equation}
\widehat{\bm{a}}=\bm{B}^{-1}\bm{Y}^{\dagger}\bm{p},
\label{eq:hoa_estimation_final}
\end{equation}
where $\bm{Y}^{\dagger}$ is the pseudoinverse of $\bm{Y}$. 
In the practical calculation, the Tikhonov regularization is introduced with the small regularization parameter, say, $10^{-4}$.

The obtained coefficients with $n=0$ and $n=1$ correspond to pressure and velocity:
\begin{equation}
p = \frac{a_{00}}{\sqrt{4\pi}},\quad
\bm{u} = \frac{1}{3i\rho_0 c}\sqrt{\frac{3}{4\pi}}
\begin{bmatrix}
a_{1x}\\ a_{1y}\\ a_{1z}
\end{bmatrix}.
\label{eq:pressure_velocity_final}
\end{equation}
From the pressure and the velocities above, the intensity and the energy can be calculated as Eq. (\ref{eqn:IandE}).
These four coefficients $\{a_{00},a_{1x},a_{1y},a_{1z}\}$ constitute the FOA components.
The higher-order coefficients are used as necessary.

For this array (64 channels, radius $r$ = 0.042 m), the spherical harmonic expansion was limited to the fixed order $N_{\mathrm{max}} = 4$.
This corresponds approximately to \(kr \approx 4\), i.e., a validity limit near \(f = \frac{4c}{2\pi r} \approx \SI{5.2}{kHz}\) for \(r=\SI{42}{mm}\).
Higher frequencies were also analyzed for completeness, but the HOA representation above this limit may include spatial aliasing effects.
The fixed-order setting was chosen to ensure a consistent comparison across arrays and to avoid overfitting the mode expansion to higher frequencies.
This 64-point Fibonacci-based equal-area distribution, modeled as a proxy for the Eigenmike, was referred to as {\bs Fibo64}, hereafter.

\subsection{Tight frame microphone array (TF24)}

A tight frame is an overcomplete basis that enables redundant yet stable reconstruction \cite{CasazzaTightFrame}. 
On the sphere, a spherical $t$-design with $t=1$ is equivalent to a tight frame \cite{HughesWaldron2021}, providing a foundation for isotropic sampling in spherical processing. Recent work extends this concept to spherical framelets \cite{XiaoZhuang2023}, optimal finite-dimensional frames\cite{BargTwoDistance}.
Building on these, the present study implements uniform directional sampling based on a tight-frame array for diffuseness and intensity measurement.

In practice, we propose a new array using multiple directional microphones arranged isotropically on a sphere, based on a tight-frame distribution, referred to as {\bs TF24}. A finite set of such points uniformly approximates the sphere while maintaining reconstruction stability.
This microphone configuration was initially devised for measuring directional impulse responses intended for sound field reproduction\cite{Omoto2020}. 
Preliminary trials were conducted to estimate acoustic intensity using this setup, which subsequently led to the more detailed investigation presented in this study.

Advantages of the proposed array include:
\begin{itemize}
\item Reduced angular-dependent errors through uniform sampling
\item Wideband operation with fewer channels
\item Easier calibration using well-characterized microphones
\item Possibility of sequential measurements by rotating a single microphone, in the case of impulse response measurement
\end{itemize}

Let the measured directional quantities, say, intensities, form $\bm{I}_{\text{meas}(N)}$, and let $\bm{R}$ be the $N\times3$ matrix of unit direction vectors $\bm{r}_i$:
\begin{equation}
\bm{I}_{\text{meas}(N)} =
\begin{bmatrix}I_{r_1}\\I_{r_2}\\\vdots\\I_{r_N}\end{bmatrix},\quad
\bm{R} =
\begin{bmatrix}
r_{1,x}&r_{1,y}&r_{1,z}\\
r_{2,x}&r_{2,y}&r_{2,z}\\
\vdots &\vdots &\vdots\\
r_{N,x}&r_{N,y}&r_{N,z}
\end{bmatrix}.
\label{eqn:ImeasN}
\end{equation}

The Cartesian intensity estimate is
\begin{equation}
\bm{I}_{\text{meas}(xyz)}=\bm{R}^{\!\top}\bm{I}_{\text{meas}(N)}.
\label{eqn:mx}
\end{equation}
For the true intensity $\bm{I}_{(xyz)}$, its projection onto $N$ measured directions is
\begin{equation}
\bm{I}_{(N)}=\bm{R}\bm{I}_{(xyz)},
\label{eqn:ItrueN}
\end{equation}
and reconstruction via the pseudoinverse yields
\begin{equation}
\bm{I}_{(xyz)}=\bm{R}^\dag\bm{I}_{(N)}=(\bm{R}^{\!\top}\bm{R})^{-1}\bm{R}^{\!\top}\bm{I}_{(N)}.
\label{eqn:Itruexyz}
\end{equation}
For a tight frame, $\bm{R}^{\!\top}\bm{R}=A\bm{I}$ is assured and this relationship gives
\begin{equation}
\bm{I}_{(xyz)}=\frac{1}{A}\bm{R}^{\!\top}\bm{I}_{(N)}.
\label{eqn:Itruexyz2}
\end{equation}
Thus, the projection sum in Eq.~(\ref{eqn:mx}) yields a scaled Cartesian intensity vector.

\medskip\noindent
\textbf{Microphones and configuration.}

Twelve directional axes are defined using twenty-four microphones at three zenith angles ($\theta=45^\circ, \; 90^\circ, \; 135^\circ$) and eight azimuths separated by $45^\circ$.  
Each opposing pair forms one velocity axis.  
The direction matrix $\bm{R}$ is
\begin{equation}
\bm{R} =
\begin{bmatrix}
1/\sqrt{2} & 0 & 1/\sqrt{2} \\
1/2 & 1/2 & 1/\sqrt{2} \\
0 & 1/\sqrt{2} & 1/\sqrt{2} \\
-1/2 & 1/2 & 1/\sqrt{2} \\
-1/\sqrt{2} & 0 & 1/\sqrt{2} \\
-1/2 & -1/2 & 1/\sqrt{2} \\
0 & -1/\sqrt{2} & 1/\sqrt{2} \\
1/2 & -1/2 & 1/\sqrt{2} \\
1 & 0 & 0 \\
1/\sqrt{2} & 1/\sqrt{2} & 0 \\
0 & 1 & 0 \\
-1/\sqrt{2} & 1/\sqrt{2} & 0
\end{bmatrix},
\label{eqn:TF24}
\end{equation}
and $\bm{R}^{\!\top}\bm{R}=4\bm{I}$ in this case.

As descrived above, the directional configuration of the Tight Frame array, based on vectors separated by 45$^\circ$ in azimuth and elevation, was originally devised by the authors as a practical microphone geometry for a hedgehog-type array intended for sound-field reproduction\cite{Omoto2020}.
Microphones oriented exactly toward the zenith and the nadir were intentionally omitted, primarily for reasons of mechanical simplicity and ease of installation and handling. As a result, the adopted configuration exhibited more stable numerical properties.

For example, if the zenith and nadir directions are additionally included, resulting in a 14×3 directional matrix in Eq. (\ref{eqn:TF24}), the matrix product $\bm{R}^{\!\top}\bm{R}$ no longer reduces to a scalar multiple of the identity matrix. In such a case, the physical quantities cannot be consistently transformed into the Cartesian $x$, $y$, and $z$ components using Eq. (\ref{eqn:mx}), thereby undermining the tight-frame property required for stable processing.

\medskip\noindent
\textbf{Directional intensity, energy, and diffuseness}

Microphone pairs aligned along twelve directions compute \emph{pseudo-}intensity via the cardioid-pair (C-C) principle~\cite{Hoshi2018,Hanyu2024}.
% here adapted for practical cardioid microphones as the \emph{pseudo} C-C method.
When measured microphone directivities are used, the corresponding quantities are distinguished by the prefix “pseudo-”, hereafter.

\begin{figure}[t]
  \includegraphics[width=0.5\linewidth,pagebox=artbox]{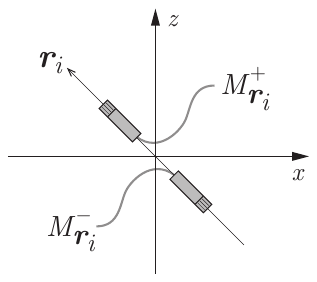}
  \caption{Schematic diagram of the signals used to construct the C-C method in the {TF24} system. 
  Twelve directional pairs are formed from 24 microphones.}
  \label{fig:mic_signals}
\end{figure}

As illustrated in Fig.~\ref{fig:mic_signals}, let $M^+_{\bm{r}_i}$ and $M^-_{\bm{r}_i}$ denote the complex amplitudes of the signals at a given frequency measured by the microphones located at the positive and negative ends of the $\bm{r}_i$ axis, respectively, for $i=1,\ldots,12$.  

The directional response of each microphone, $d(\theta)$, where $\theta$ denotes the angle from its main axis, 
is approximated by a polynomial of $\cos\theta$ as
\begin{equation}
d(\theta) = \sum_{n=0}^{N} a_n \cos^n \theta.
\end{equation}
The case of $N=1$ with $a_0 = 0.5$ and $a_1 = 0.5$ corresponds to an ideal cardioid pattern.

When two microphones are aligned along the same axis but oriented in opposite directions, 
their directional responses can be expressed as $d(\theta)$ and $d(\pi - \theta)$, respectively.  
The sum and difference of their complex amplitudes, $ M^+_{\bm{r}_i} \pm M^-_{\bm{r}_i}$, then form symmetric and antisymmetric combinations of these directional responses.  
%The sum component primarily exhibits an omnidirectional characteristic, 
%approximating the local sound pressure, whereas the difference component resembles a figure-of-eight pattern, 
%corresponding to the particle velocity component along $\bm{r}_i$, provided the directional characteristics is ideal cardioid.
%Hence, the pairwise sum and difference of oppositely oriented practical (not ideal) cardioid microphones provide an operational means 
%of approximating the local pressure and the corresponding particle-velocity component.
For an ideal cardioid directivity, the symmetric (sum) component exhibits an omnidirectional response and thus provides an estimate of the local sound pressure.
Conversely, the antisymmetric (difference) component approaches a figure-of-eight directivity aligned with $\bm{r}_i$, corresponding to the particle-velocity component along that axis.

Although practical cardioid microphones deviate from this ideal behavior, the pairwise sum and difference of oppositely oriented directional microphones still offer an effective operational approximation of the local pressure and velocity.

The resulting quantities are defined as
\begin{align} 
\widehat{p}_{\bm{r}_i} & = M^+_{\bm{r}_i} + M^-_{\bm{r}_i}, \label{eqn:p}\\ 
\widehat{u}_{\bm{r}_i} &= \frac{M^+_{\bm{r}_i} - M^-_{\bm{r}_i}}{Z_0}, \label{eqn:u}\\
\widehat{I}_{\bm{r}_i}&= \frac{|M^+_{\bm{r}_i}|^2 - |M^-_{\bm{r}_i}|^2}{2 Z_0}, \label{eqn:I}\\
\widehat{E}_{\bm{r}_i} & = \frac{|p_{\bm{r}_i}|^2}{4 \rho_0 c^2}  + \frac{\rho_0}{4} |u_{\bm{r}_i}|^2. \label{eqn:E}
\end{align}
Since the directional elements used in the array do not exhibit strictly cardioid directivity; their directivity varies with frequency.
Therefore, the \emph{pseudo-}pressure $\widehat{p}_{\bm{r}_i}$ obtained from their summation and the \emph{pseudo-}velocity $\widehat{u}_{\bm{r}_i}$ inferred from their differential output cannot be regarded as the exact physical quantities.
Instead, they represent {\bf operational} approximations of the local sound pressure and particle velocity, derived under the assumption of complementary directivity pairs.
The quantity corresponding to the intensity, $\widehat{I}_{\bm{r}_i}$, and the energy density, $\widehat{E}_{\bm{r}_i}$, are thus defined in this approximate sense and referred to here as \emph{pseudo-}intensity and \emph{pseudo-}energy density.
Representative examples of these directivity patterns are presented in the next section.

Similarly, the directional dependent diffuseness index based on the I/E concept is defined as
\begin{equation}
\Psi_{\bm{r}_i} = 1 - \frac{| \widehat{I}_{\bm{r}_i} |}{\: c \widehat{E}_{\bm{r}_i}\:},
\label{eqn:psi_r}
\end{equation}
and its weighted extension is discussed in the following section.

%%%%%%%%%%%%%%%%%%%%%%%%%%%%%%%%%%%%%%%%%

\section{Benchmark problems for numerical simulation}
\label{sec:benchmark}

Unlike previous studies that evaluated diffuseness primarily within the spherical harmonic domain under ideal conditions, the present work reformulates the covariance analysis in the first-order velocity space and verifies its performance with physically realizable microphone geometries. By integrating Epain and Jin’s benchmark paradigm\cite{Epain2016} with a tight-frame sampling approach, this study provides a unified yet experimentally grounded framework for diffuseness characterization.

\subsection{Overview of simulation framework}

This section defines benchmark problems to compare the diffuseness-evaluation methods for the three array configurations introduced in the previous section ({\bs Afmt}, {\bs Fibo64}, and {\bs TF24}). Each benchmark specifies a virtual acoustic scene with known structure; from the simulated microphone pressures, we compute diffuseness and assess accuracy, stability, and frequency dependence.

\begin{figure}[t]  % r: right, l: left
  \includegraphics[width=0.75\linewidth,pagebox = artbox]{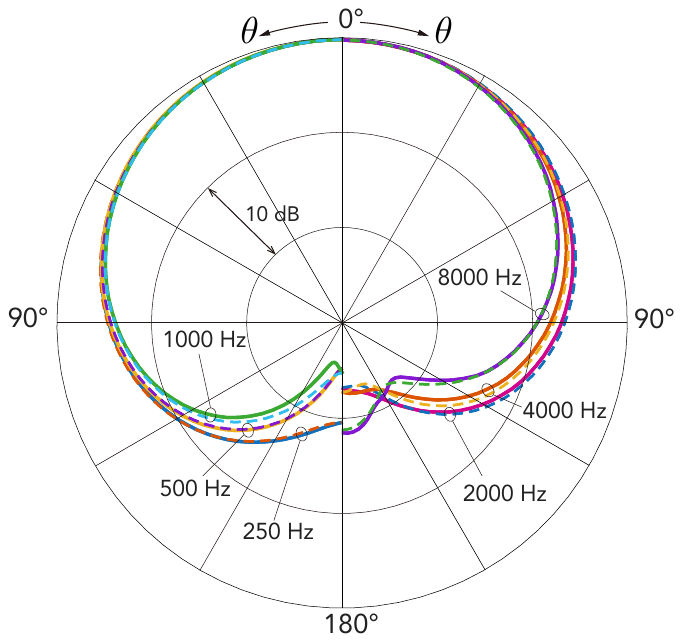}
  \caption{Measured (solid lines) and modeled (broken lines) directivities of the DPA2012 cardioid microphone assumed in the {\bs Afmt} and the {\bs TF24} array. To avoid visual clutter, the polar plots show the responses for 250 Hz to 1 kHz on the left half and for 2 kHz to 8 kHz on the right half. In the calculations presented in the text, however, a wider frequency range from 63 Hz to 16 kHz was analyzed in octave bands.}
  \label{fig:directivities}
\end{figure}

The {\bs Afmt} array was modeled after the AMBEO VR Mic, and the capsule directivities were approximated by a cardioid model consistent with DPA 2012. 
This microphone is also used for the {\bs TF24} array. The directivity patterns were measured in an anechoic chamber at $3^{\circ}$ resolution and fitted as a polynomial in $\cos\theta$ (power-series form) up to 8th order (as shown in Fig.~\ref{fig:directivities}); these fits were used in simulation.

\begin{figure}[t] 
  \includegraphics[width=\linewidth,pagebox = artbox]{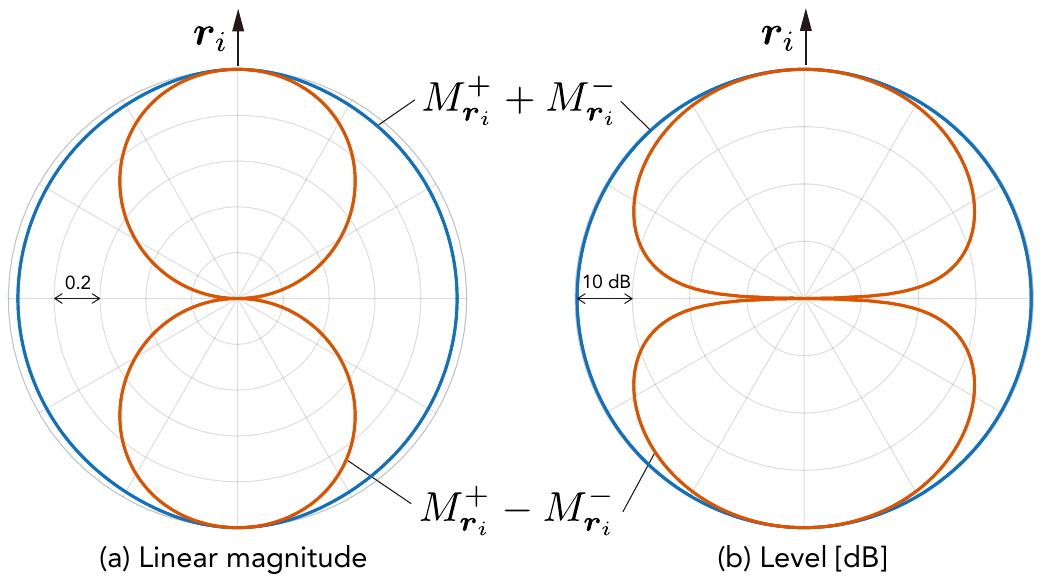}
\caption{Representative example of modeled directional responses derived from the microphone directivity polynomial in Eq. (\ref{eq:directivity_model}) at 1 kHz.
The results are shown (a) in linear magnitude and (b) in level (dB).
The outer curve represents the pseudo-omnidirectional response obtained from the sum signal, whereas the inner curve corresponds to the pseudo-figure-of-eight response obtained from the difference signal. Both responses were derived from a pair of oppositely oriented microphones placed in the TF24 array with an inter-microphone spacing of 20 mm.}
\label{fig:sum_and_sub}
\end{figure}

Array geometries were set as follows (see Fig.~\ref{fig:mic_directions}): the A-format capsules are located 6~mm from the origin [Fig.~\ref{fig:mic_directions}(a)]; the rigid-sphere array uses a radius of 42~mm [Fig.~\ref{fig:mic_directions}(b)].
The spacing between microphones in the {\bs TF24} array was set to 20\,mm; that is, the distance from the origin in [Fig.~\ref{fig:mic_directions}(c)] is 10\,mm. 
Although this dimension may seem unrealistic for practical microphones approximately 100\,mm in length, 
it represents a practical configuration assuming sequential impulse-response measurements by mechanically rotating a single microphone.

Simulations are performed in the frequency domain. Octave bands from \SI{63}{Hz} to \SI{16}{kHz} were analyzed.
Sources are either a single plane wave or a sum of multiple plane waves. 
To emulate band-limited noise, we draw 100 frequencies uniformly in log scale within a one-octave band and sum sinusoids with random amplitude and phase.

\medskip\noindent
\textbf{Pressure at each microphone}

The complex pressure at the $m$-th microphone for frequency $f$ (wavenumber $k=2\pi f/c$) due to $N_s$ plane waves incident from directions $\hat{\nu}_i$ is expressed as
\begin{equation}
p_m(f)
= S_m(f)
  \sum_{i=1}^{N_s}
  D_m\!\bigl(\hat{\nu}_i,f;\hat{d}_m\bigr)
  A_i\, e^{\,j k\,\bm{r}_m \cdot \hat{\nu}_i + j\varphi_i},
\label{eq:pm_directivity}
\end{equation}
where $\bm{r}_m$ is the position vector of the $m$-th microphone,
$\hat{\nu}_i$ is the unit vector of the $i$-th plane-wave propagation direction,
$A_i$ and $\varphi_i$ are its amplitude and phase, respectively,
$S_m(f)$ denotes the on-axis sensitivity of the microphone, and
$D_m(\hat{\nu}_i,f;\hat{d}_m)$ represents the directional response of the microphone whose principal axis points along the unit vector $\hat{d}_m$.
The exponential term $e^{j k \bm{r}_m\!\cdot \hat{\nu}_i}$ accounts for the phase delay due to the microphone position.

For the present simulations, the directivity pattern was modeled as a polynomial function of $\cos\theta$, where
$\theta$ is the angle between $\hat{d}_m$ and $\hat{\nu}$:
\begin{equation}
D_m(\hat{\nu},f;\hat{d}_m)
\simeq
\sum_{n=0}^{8} a_n(f)\cos^n\theta,
\qquad
\theta=\arccos(\hat{d}_m\!\cdot\!\hat{\nu}).
\label{eq:directivity_model}
\end{equation}
This fitted model corresponds to the broken lines shown in Fig.~\ref{fig:directivities}.

As an example, Fig.~\ref{fig:sum_and_sub} illustrates the modeled directivities obtained from the sum and difference of opposite microphone pairs
based on the polynomial model in Eq.~(\ref{eq:directivity_model}) at 1 kHz.
The sum response approximates a pseudo-omnidirectional pattern, while the difference response approximates a pseudo-figure-of-eight pattern.

For the {\bs Fibo64} rigid-sphere model with omnidirectional elements, $D_m=1$; directivity/baffle effects are captured by the radial filters $b_n(ka)$ in the Spherical Harmonic expansion. 
Thus, after computing $p_m(f)$ via \eqref{eq:pm_directivity}, HOA coefficients are estimated from the SH model as described in Sec.~II.

\begin{figure}[t] 
  \centering
  \includegraphics[width=\linewidth,pagebox=artbox]{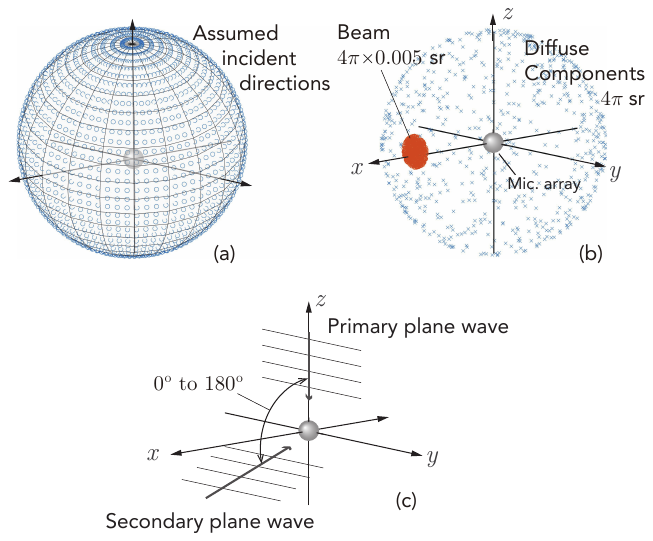}
 \caption{Conceptual diagrams of the incident directions of plane waves used in the benchmark simulations:
(a) single plane-wave incidence from various directions;
(b) diffuse components arriving randomly over the entire solid angle of $4\pi$, and beam components arriving within a confined solid angle corresponding to 0.5 \% of the total sphere. 
The center of the beam was set at 3° in azimuth and 87° in zenith angles.
(c) An interference field formed by two plane waves. One (the primary wave) was incident from the zenith direction ($\theta$ = 0°), while the other secondary wave was incident from variable zenith angles ranging from 0° to 180°, producing an interference pattern used to evaluate the diffuseness under such conditions.}
  \label{fig:beam}
\end{figure}

\subsection{Benchmark case definitions}

As shown in Fig. \ref{fig:beam}, three representative cases are assumed for numerical simulation.

\subsubsection*{Case~1: Single plane-wave incidence.}  
A single plane wave arrives from a given direction; the theoretical diffuseness is $\Psi=0$. We sweep both elevation and azimuth over the sphere to test directional robustness and verify how accurately intensity-based direction-of-arrival can be recovered. 
For Case 1, 72 azimuthal directions (0$^\circ$ to 355$^\circ$) and 35 zenith angles (5$^\circ$ to 175$^\circ$), both at 5$^\circ$ intervals, were simulated, resulting in 2520 incidence directions in total.
For each case, the mean diffuseness and the error in arrival-angle detection based on the pseudo-intensity were evaluated to assess the accuracy of sound-arrival direction estimation for each array.

\subsubsection*{Case~2: Plane-wave beam + diffuse mixture.}  
As illustrated in Fig.~\ref{fig:beam}, we mix a beam arriving from a narrow solid angle (here $0.5\%$ of $4\pi$ sr) with diffuse components uniformly distributed over $4\pi$ sr. We define the beam-to-total energy ratio
\begin{equation}
\eta = \frac{E_{\mathrm{beam}}}{E_{\mathrm{beam}} + E_{\mathrm{diffuse}}},
\end{equation}
where $E_{\mathrm{beam}}$ is the energy of beam, and $E_{\mathrm{diffuse}}$ is the energy of diffuse components. The $\eta$ was varied from 0 to 1. Both the diffuse and beam components use 100,000 rays with random amplitude and phase; for band-limited noise, we draw 100 log-spaced frequencies within a 1-octave band at each trial. We examine whether the estimated $\Psi$ is stable and how linearly it tracks $\eta$.

Case 1 and Case 2 are used not only as benchmark scenarios for diffuseness estimation but also as controlled tests to quantify the deviation between the proposed pseudo quantities and the ideal co-located pressure-velocity (C-C) reference.

\subsubsection*{Case~3: Reactive field (interference).}  

In this case, an interference field was simulated by superposing two plane waves: a primary wave incident from the zenith direction and a secondary wave arriving from variable zenith angles between 0° and 180°. For each configuration, 1,000 realizations were generated by randomly varying the phase (within $\pm 2 \pi$) and amplitude (within $\pm 3$ dB) of the secondary wave. The pseudo-intensity, pseudo-energy density, and covariance matrix were computed for each realization, and the final values were obtained by averaging these quantities over all realizations to represent the mean behavior of the interference field.

\subsection{Evaluation metrics}

Given the array pressures, the following diffuseness indices are calculated.

\subsubsection*{Intensity-Energy ratio $\Psi_{\mathrm{IE}}, \: \Psi_{\mathrm{AVE}}$, Coefficient-of-variation $\Psi_{\mathrm{CV}}$}  

From (pseudo) pressure and (pseudo) velocity (via A-format or HOA/FOA processing) in the cases of {\bs Afmt} and {\bs Fibo64}, the (pseudo) intensity $\bm{I}(f)$ and the (pseudo) energy density $E(f)$ are computed and the diffuseness $\Psi_{\mathrm{IE}}$ is defined as 
\begin{equation}
\Psi_{\mathrm{IE}}(f) = 1 - \frac{|\bm{I}(f)|}{c\,E(f)}.
\label{eq:IE}
\end{equation}

$\Psi_{\mathrm{IE}}\!\approx\!0$ indicates a traveling-wave field; $\Psi_{\mathrm{IE}}\!\approx\!1$ indicates a diffuse field. 

For {\bs TF24} case, we compute directional $\Psi_{r_i}$ defined as Eq. (\ref{eqn:psi_r}). 
%In directions where only uncorrelated background noise is observed, the estimated diffuseness tends to assume relatively high values. However, such regions do not correspond to acoustically significant signals. This observation suggests the necessity of introducing directional weighting. 
In directions where the incident sound energy is negligible, the pseudo pressure and particle-velocity components estimated by the TF24 array become very small and are comparable to background or numerical noise.
Because the diffuseness measure is defined as a normalized ratio, the residual uncorrelated components can become relatively dominant in such low-energy directions, leading to artificially high diffuseness values.
Importantly, these regions do not correspond to acoustically meaningful sound-field contributions, but rather to directions with insufficient signal energy.
This behavior indicates that directional weighting based on signal strength is necessary to suppress physically irrelevant directions when evaluating diffuseness.

\begin{figure}[t] 
  \centering
  \includegraphics[width=0.9\linewidth,pagebox=artbox]{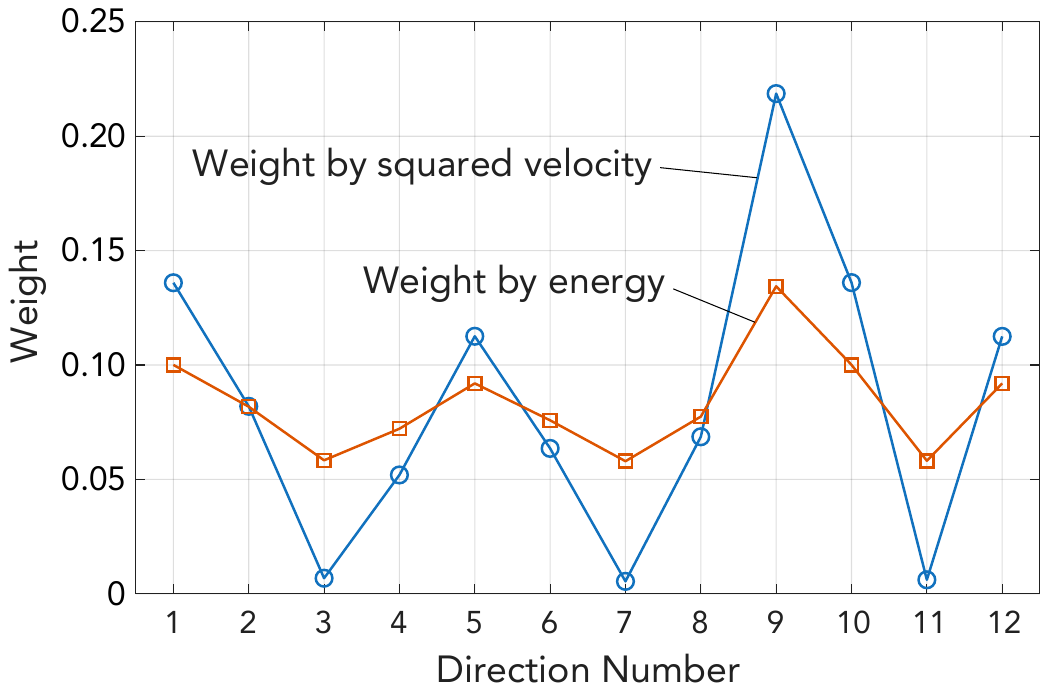}
 \caption{Numerical simulation results of directional weighting applied to the diffuseness estimation.
Results are shown for Case 2 with a beam ratio of $\eta = 0.95$ at 2 kHz.
In this condition, 95\% of the acoustic energy is carried by a single plane wave arriving approximately from the $x$-axis direction.
This dominant incidence corresponds to the 9th row of the direction vector matrix $\bm{R}$ in Eq. (\ref{eqn:TF24}), and therefore the squared particle-velocity (or energy) associated with the 9th direction becomes significantly larger than those of the other directions.}
  \label{fig:weights}
\end{figure}

After several preliminary trials, weighting by the squared magnitude of the particle velocity was adopted, as it exhibited higher sensitivity to spatial variations in the sound field. 
Fig. \ref{fig:weights} illustrates an example of the directional energy distribution and the corresponding weighting factors based on the squared particle velocity, computed for Case 2. 
The incident beam is assumed to arrive predominantly from the 9th direction (9th row of the direction vector matrix $\bm{R}$ in Eq. (\ref{eqn:TF24}), i.e., $x$-axis), and it can be observed that the velocity-based weighting responds subtly but consistently to this directional bias.

\begin{equation}
\Psi_\mathrm{AVE}= \displaystyle
\frac{\displaystyle \sum_{i=1}^{12}\Psi_{\bm{r}_i} \: |u_{\bm{r}_i}|^2 }
{\displaystyle \sum_{i=1}^{12} \: | u_{\bm{r}_i} | ^2}.
\label{eqn:psi_ave}
\end{equation}

The appropriateness of this weighting scheme remains a subject for further investigation; nevertheless, several representative examples are presented below.
The possible disadvantages associated with this choice are also discussed in the Discussion section.

Finally, in addition to the intensity-energy ratio based measures,
we also consider a coefficient-of-variation-based diffuseness indicator,
denoted as $\Psi_{\mathrm{CV}}$.

The coefficient-of-variation approach was originally introduced by Ahonen\cite{Ahonen2009} and later employed in the context of spatial audio by Del Galdo \textit{et al.}\cite{DelGaldo2012}, where diffuseness is characterized by the statistical variability of acoustic intensity.
In this study, $\Psi_{\mathrm{CV}}$ is formulated exclusively in terms of the acoustic intensity vector, without explicit use of the energy density.

For a set of realizations (or time--frequency samples), $\Psi_{\mathrm{CV}}$ is defined as
\begin{equation}
\Psi_{\mathrm{CV}} =
\sqrt{1 - \frac{\|\mathbb{E}[\bm{I}(f)]\|}{\mathbb{E}[\|\bm{I}(f)\|]}},
\label{eq:Psi_CV}
\end{equation}
where $\bm{I}$ denotes the instantaneous acoustic intensity vector and $\mathbb{E}[\cdot]$ represents expectation over realizations.
This formulation yields $\Psi_{\mathrm{CV}}\approx 0$ for a highly directional (traveling-wave) field and $\Psi_{\mathrm{CV}}\approx 1$ for an isotropic diffuse field.

Unlike $\Psi_{\mathrm{IE}}$ and $\Psi_{\mathrm{AVE}}$, which rely on the ratio between mean intensity and energy density,
$\Psi_{\mathrm{CV}}$ quantifies diffuseness purely through the statistical dispersion of the intensity vector.
In the present work, $\Psi_{\mathrm{CV}}$ is included as a reference intensity-based diffuseness measure, enabling direct comparison with the proposed directionally weighted diffuseness metrics and their behavior is specifically examined using benchmark Case~2 mentioned above.

\subsubsection*{Eigenvalue-based diffuseness ($\Psi$\textsubscript{PR}, $\Psi$\textsubscript{COM})}  

To capture spatial uniformity, covariance-based measures are employed. 
In conventional approaches, a 4×4 covariance matrix is constructed from the sound pressure and three velocity components, i.e., FOA components as 
\[
\bm{C}_{\mathrm{pu}} = \langle [p,\,u_x,\,u_y,\,u_z]^H [p,\,u_x,\,u_y,\,u_z] \rangle,
\]
where $u_x, \; u_y, \; u_z$ are the $x, y, z$ directional components of velocity, and $\bm{C}_\mathrm{pu}$ represents the full energy coupling between pressure and particle velocity. 

In the present study, however, we emphasize a velocity (including pseudo) only route and form
\begin{equation}
\bm{C}_{\mathrm{uu}} = \langle \bm{u}(f)\,\bm{u}^H(f)\rangle,
\label{eq:CovU}
\end{equation}
which directly captures the spatial anisotropy of the local (pseudo)velocity distribution without an explicit reference to the pressure term.

This choice is first motivated by the characteristics of the Tight frame array. 
Since each microphone in the {\bs TF24} configuration uses a model of practical directional characteristics rather than an ideal cardioid, the sum of the outputs of each microphone pair does not necessarily reproduce the ideal omnidirectional pressure component. 

Consequently, defining diffuseness in terms of pressure-velocity coherence may become less reliable in {\bs TF24}. 
Instead, by constructing the covariance matrix solely from the velocity vectors, we can evaluate the spatial uniformity in a physically consistent and array-independent manner, 
thus ensuring the robustness of diffuseness estimation even under non-ideal directivity conditions.

In the present study, the covariance matrix was constructed solely from the velocity components derived from the FOA signals, not only for the{\bs TF24} system but also for the {\bs Afmt} and {\bs Fibo64} configurations. 
A further dominant motivation for this choice lies in the scaling issues that inevitably arise when handling both pressure and velocity simultaneously. When estimating diffuseness from eigenvalues, it is desirable to ensure not only amplitude scaling, such as SN3D normalization (e.g., Eq. (\ref{eq:pressure_velocity_final})) but also detailed frequency response consistency. 
Achieving this would require measurements in a sound field that can be regarded as \emph{sufficiently diffuse} and the computation of a corresponding whitening matrix. To avoid this complexity, the present work primarily adopts an approach based on velocity-only metrics.

In the {\bs TF24} system, the pseudo-velocity components used to construct the covariance matrix $\bm{C}_{uu}$ are obtained by projecting the vectors $u_{\bm{r}_i}$, 
measured in 12 directions, onto the global $x,\; y$, and $z$ axes. This projection is performed by multiplying each directional component by the corresponding direction vector $\bm{R}$ as,
\begin{equation}
\bm{u} = 
\begin{bmatrix}
u_x \\
u_y \\
u_z
\end{bmatrix} \approx \frac{1}{4}
\bm{R}^{\! \top} \widehat{\bm{u}}_{\bm{r}},
\end{equation}
where $\widehat{\bm{u}}_{\bm{r}} = \left [ \widehat{u}_{\bm{r}_1} \widehat{u}_{\bm{r}_2} \cdots \widehat{u}_{\bm{r}_{12}} \right ]^{\! \top}$.

Let the eigenvalues of Eq.~(\ref{eq:CovU}) be $\lambda_1 \ge \lambda_2 \ge \lambda_3$. 
From these eigenvalues, several indices can be derived. In this paper, the following two diffuseness indices are calculated and examined:

\begin{itemize}
\item \textbf{Participation Ratio  ($\Psi$\textsubscript{PR})}  
Quantifies the effective number of eigenmodes contributing to the covariance structure:
\begin{equation}
\Psi_{\mathrm{PR}} = \frac{\displaystyle \left( \sum_{i=1}^{3}\lambda_i \right )^2}{\displaystyle 3\sum_{i=1}^{3}\lambda_i^2}.
\end{equation}
This index is sensitive to how evenly the energy is distributed among the eigenvalues, 
taking a value close to $1/3$ for a single dominant mode and approaching 1 for a uniform distribution.
Note that \(\Psi_{\mathrm{PR}}\in[1/3,\,1]\) for three non-negative eigenvalues.

\item \textbf{COMEDIE ($\Psi$\textsubscript{COM})}  
Proposed by Epain and Jin, this measure normalizes the standard deviation of the eigenvalues\cite{Epain2016}:
\begin{equation}
\Psi_{\mathrm{COM}} = 1 - \sqrt{\frac{3}{2}}\,\frac{\sigma_\lambda}{\overline{\lambda}},
\end{equation}
where $\sigma_\lambda$ and $\overline{\lambda}$ denote the standard deviation and mean of $\{\lambda_i\}$, respectively.
This provides a statistically stable and physically interpretable indicator of isotropy.
\end{itemize}

%%%%%%%%%%%%%%%%%%%%%%%%%%%%%%%%%%%%%%%%%

\section{Results and discussion}
\label{sec:results}

\subsection*{Deviation of pseudo quantities from the C-C reference}

Before presenting the main simulation results, the validity of cardioid microphone modeling and the subsequent estimation of pseudo-pressure and pseudo-velocity using the C-C method are first examined.
This validation is conducted using Case 1, which serves as a baseline problem for assessing the robustness of the proposed framework.

In this study, the directional responses of practical cardioid microphones are modeled and incorporated into a {\bs TF24} configuration.
The pseudo-pressure and pseudo-velocity signals are then derived from paired microphone outputs using the C-C method.
The adequacy of this modeling and processing chain is evaluated by comparing the resulting pseudo-intensity estimates with those obtained under idealized conditions.

Specifically, two configurations are considered:
\begin{enumerate}[label = \roman*)]
\item an array composed of ideal cardioid directivities, defined as $(1-\cos\theta)/2$, and
\item an array using measured cardioid directivities, which include realistic deviations in amplitude, phase, and orientation.
\end{enumerate}
For both configurations, the accuracy of direction estimation by pseudo-intensity is evaluated, together with the ratio between the pseudo-intensity and pseudo-energy quantities (I/E).

The latter metric is introduced as a combined indicator that reflects not only angular estimation errors but also quantitative deviations in the reconstructed intensity magnitude.
As such, it provides a more comprehensive measure of the impact of microphone directivity imperfections than angular error alone.

To systematically investigate these effects, four levels of perturbations, denoted as {\bs L0} - {\bs L3}, are considered for the measured cardioid data.
These levels represent increasing degrees of phase, amplitude, and angular deviations, corresponding to plausible and conservative ranges of manufacturing and calibration uncertainties.
For each level, the deviations are modeled as zero-mean random perturbations characterized by the standard deviations listed in Table~\ref{tab:deviation_levels}, and the resulting quantities are evaluated using 200 Monte Carlo realizations.
The specific definitions of {\bs L0} - {\bs L3} are summarized in Table~\ref{tab:deviation_levels}.

\begin{table}[t]
\centering
\caption{Assumed levels of realistic deviations in cardioid directivity used in Case 1. The values indicate standard deviations of the imposed perturbations, and each condition is evaluated using 200 Monte Carlo trials.}
\label{tab:deviation_levels}
%\begin{tabular}{|*{4}{>{\centering\arraybackslash}p{0.18\textwidth}|}}
\begin{tabular}{cccc}
\hline
Level & Gain [dB] & Phase [deg] & Axis-misalign. [deg] \\
\hline
{\bs L0} & 0.0 & 0  & 0 \\
{\bs L1} & 0.5 & 5  & 1 \\
{\bs L2} & 1.0 & 10 & 3 \\
{\bs L3} & 2.0 & 20 & 5 \\
\hline
\end{tabular}
\end{table}

\begin{figure}[t] 
  \centering
  \includegraphics[width=0.9\linewidth,pagebox=artbox]{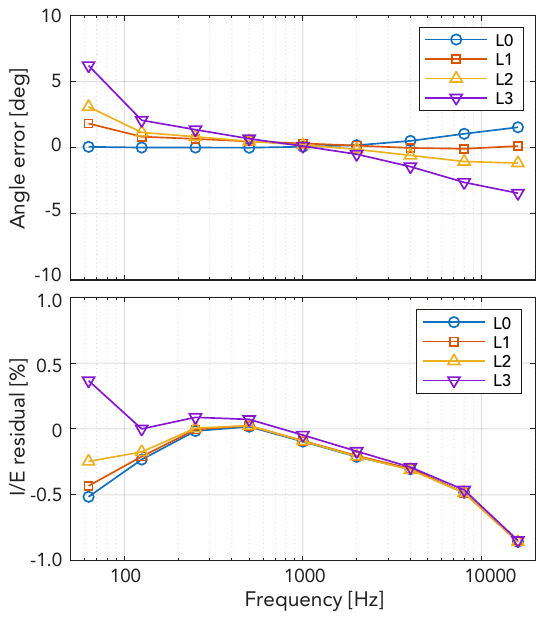}
 \caption{Numerical simulation results of frequency-dependent performance degradation caused by realistic deviations in cardioid directivity (Table \ref{tab:deviation_levels}), evaluated as the difference between modeled and ideal cardioid configurations.
The upper panel shows the angular error penalty, and the lower panel shows the corresponding I/E residual penalty.
Each curve represents the direction median of the Monte Carlo 90th-percentile error.}
  \label{fig:c_error}
\end{figure}

Figure \ref{fig:c_error} summarizes the validation results for Case 1, showing the frequency dependence of the performance degradation caused by realistic deviations in cardioid directivity.
The upper panel presents the angular error penalty, while the lower panel shows the corresponding I/E residual penalty, both defined as the difference between the measured and ideal cardioid configurations.

For each frequency band, the angular and I/E errors were first evaluated for all incident directions using Monte Carlo simulations.
The 90th percentile of the error distribution was computed for each direction, and the median across all directions was then used as a representative measure.
All curves, therefore, represent conservative, direction-averaged penalties relative to the ideal cardioid case.

As shown in the figure, the error-free condition {\bs L0} exhibits negligible angular deviation in the mid-frequency range, together with I/E residuals well within 1\%.
With increasing perturbation levels from {\bs L1} to {\bs L3}, the magnitude of the deviations increases, as expected.
However, even in the most severe case considered here, the absolute errors remain small across the entire frequency range, indicating that the impact of realistic cardioid imperfections is limited in practical situations.

Based on these results, the proposed modeling of practical cardioid directivities and the associated C-C-based processing are considered sufficiently robust.
In the following, simulation results incorporating measured cardioid characteristics are therefore presented without further modification.

\subsection*{Case~1: Diffuseness and sound-arrival direction estimation for a single plane wave}

\begin{figure}[t] 
  \centering
  \includegraphics[width=0.9\linewidth,pagebox=artbox]{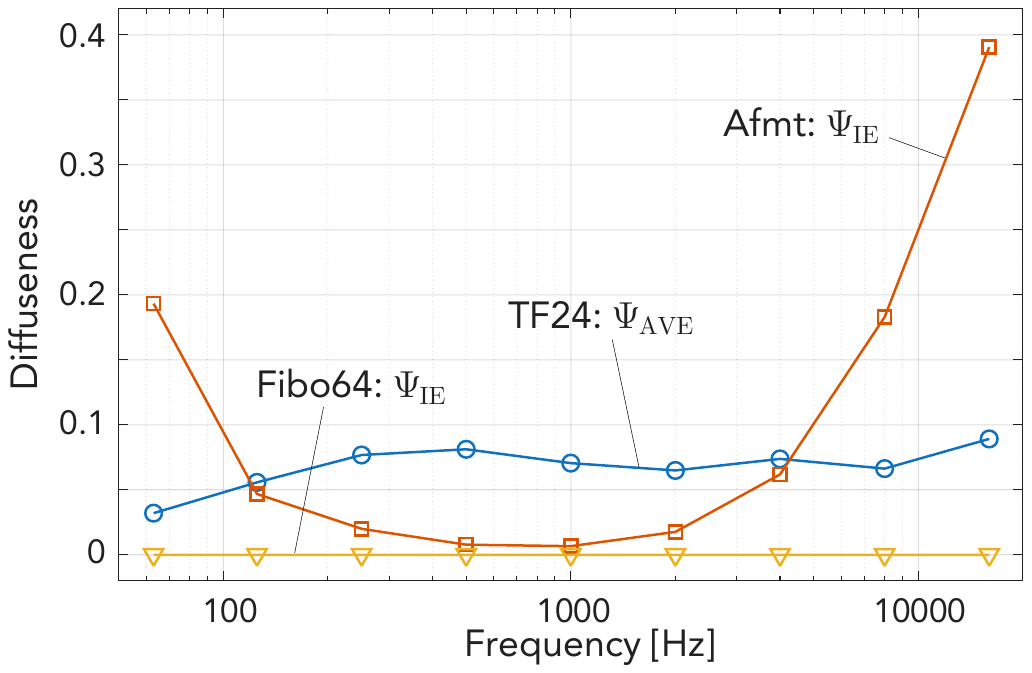}
 \caption{Case 1 (numerical simulation): Diffuseness indices based on the I/E method: 
for {\bs Afmt} and {\bs Fibo64}, the results show $\Psi_{\mathrm{IE}}$; 
for {\bs TF24}, $\Psi_{\mathrm{AVE}}$ is presented.}
  \label{fig:Psi_IE_AVE}
\end{figure}

A single plane wave was simulated to impinge from all directions, and the I/E based diffuseness indices $\Psi_{\mathrm{IE}}$ for {\bs Afmt} and {\bs Fibo64}, and $\Psi_{\mathrm{AVE}}$ for {\bs TF24}, and eigenvalue-based indices $\Psi_{\mathrm{PR}}$, and $\Psi_{\mathrm{COM}}$---were computed.  
In theory, all indices should yield $\Psi=0$, but small deviations were observed depending on the microphone array configuration, especially in the case of I/E based indices.
Fig. \ref{fig:Psi_IE_AVE} shows the results. The averaged value of diffuseness for all incident angles is shown at every frequency.

The {\bs Afmt} results exhibited relatively large values in the low-frequency region and above 4\,kHz, 
while the {\bs TF24} array also showed slightly elevated values of below 0.1 at a wide frequency range. 
In contrast, the {\bs Fibo64} array, which employs 64 microphones, yielded values close to zero across all frequency bands.

For the eigenvalue-based indices, although the absolute magnitudes differed across methods, 
all 2,520 incident directions yielded values of zero up to three decimal places 
when averaged over the nine octave bands from 63\,Hz to 16\,kHz; therefore, these results are not shown in the figure.

Under the condition of single plane-wave incidence, 
the superiority of the eigenvalue-based analysis using the covariance matrix was demonstrated. 
Among the microphone arrays, the {\bs Fibo64} configuration exhibited the most stable results.

While the upper frequency limit of the {\bs Fibo-64} array can be roughly predicted from the spatial aliasing condition, the simulation results exhibit stable performance even above this range.
This suggests that the quasi-uniform sampling of the Fibonacci lattice provides smoother angular coverage and mitigates high-order aliasing components that typically degrade array accuracy.

%%%%%%

\begin{figure}[t] 
  \centering
  \includegraphics[width=0.9\linewidth,pagebox=artbox]{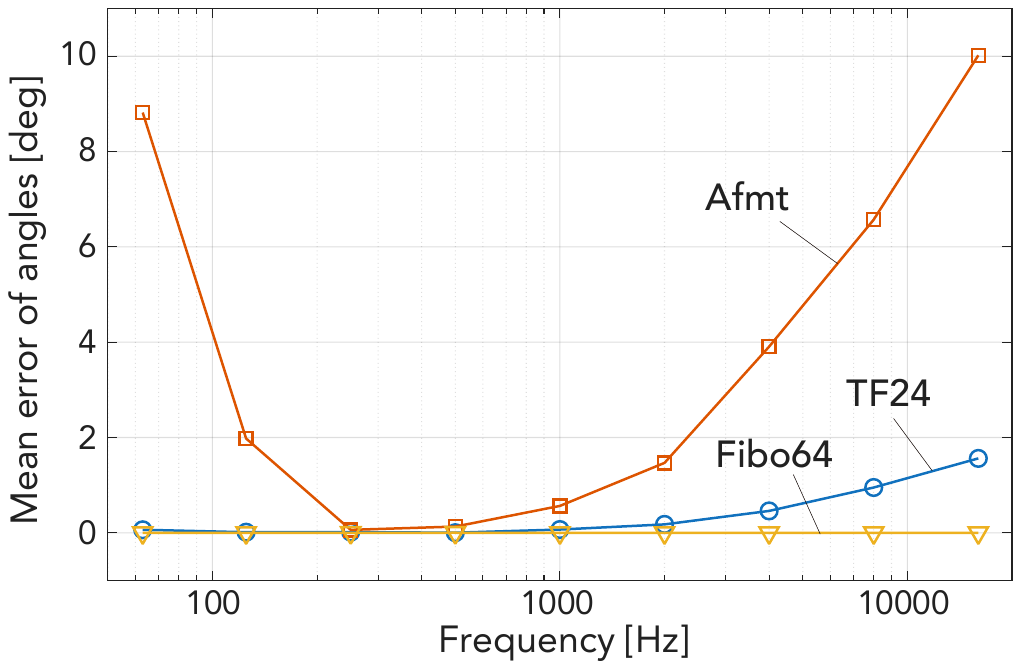}
 \caption{Case 1 (numerical simulation): Mean detection error of the incident angle estimated from acoustic intensity obtained by each microphone array:
Average detection error over 2520 incident directions.}
  \label{fig:angle_error}
\end{figure}

Directional accuracy of the active-intensity vector was also compared. The results are shown in Fig. \ref{fig:angle_error}.
Although diffuseness and the angular error in intensity detection are entirely different quantities, the overall trend of the results is similar to that shown in Fig. \ref{fig:Psi_IE_AVE}. 
For the {\bs Afmt} system, the estimation accuracy is low in both the low- and high-frequency ranges, and the frequency band in which the correct incident direction can be detected is limited to approximately 500 Hz to 2 kHz. 
In contrast, the {\bs Fibo64} system achieves almost error-free direction detection across the entire frequency range. 
The {\bs TF24} array employing cardioid microphones demonstrates angular detection accuracy comparable to that of the {\bs Fibo64} configuration.
Specifically, the error remains within 1$^\circ$ below 4 kHz, effectively matching the performance of {\bs Fibo64} in this frequency range.
Although a moderate increase in error is observed at higher frequencies, it does not exceed 2$^\circ$, suggesting that {\bs TF24} maintains robust directional performance across a wide frequency range.
Considering that these values represent the average error over 2,520 incident directions, the overall performance can be regarded as practically sufficient.

%---

\subsection*{Case~2: Linearity of diffuseness in plane-wave and diffuse mixture}

Next, a mixture of a narrow beam and an isotropic diffuse field was simulated by varying the energy ratio $\eta$ between beam and diffuse components.  

\begin{figure}[t] 
  \centering
  \includegraphics[width=0.9\linewidth,pagebox=artbox]{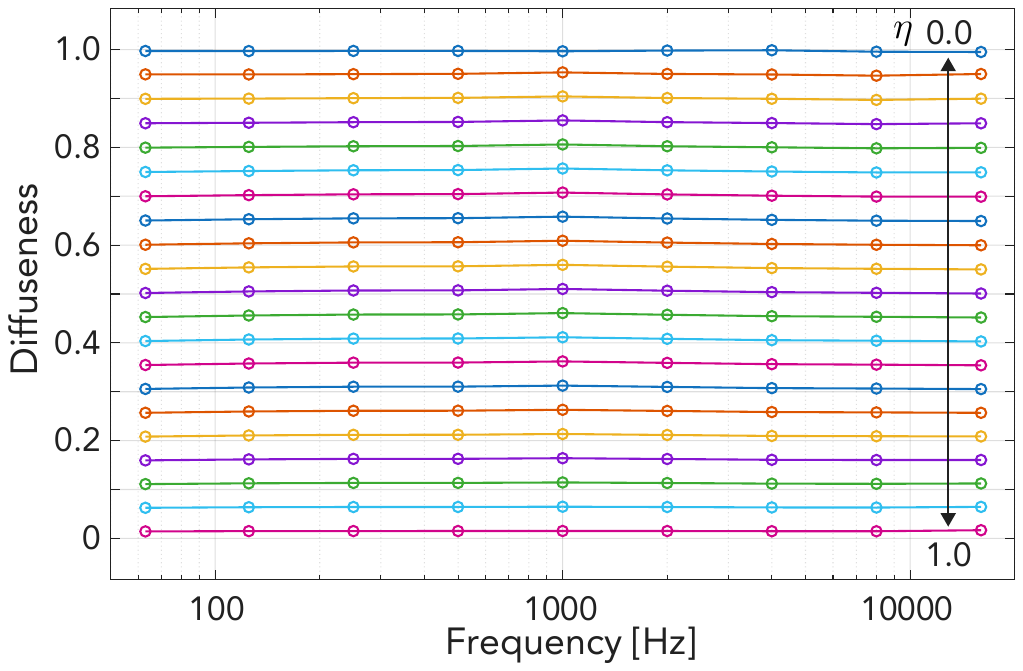}
 \caption{Case 2 (numerical simulation): Example of the variation in diffuseness when the beam ratio $\eta$ is varied from 0 to 1 in steps of 0.05.
Results are obtained for the {\bs TF24} array using the eigenvalues of the covariance matrix and the COMEDIE, i.e., $\Psi_{\mathrm{COM}}$.
The numbers in the figure indicate $\eta$; in the ideal case, the diffuseness corresponds to $1-\eta$.}
  \label{fig:psi_ratio}
\end{figure}

Figure~\ref{fig:psi_ratio} shows the results of the COMEDIE-based diffuseness index, $\Psi_{\mathrm{COM}}$, for the {\bs TF24} system. 
The results closely follow the theoretical value of $1 - \eta$ from the low to mid frequencies. 
All indices exhibit an approximately linear relationship, and the correlation between $1 - \eta$ and the diffuseness estimated in each frequency band is very high. 
Because the indices are therefore difficult to distinguish in direct comparison, we quantified their deviation more explicitly by calculating the absolute difference between the diffuseness value and $1 - \eta$, and taking the maximum of this difference for each index. 
%The results are shown in Fig.~\ref{fig:ratio_error}.

%In Fig.~\ref{fig:ratio_error}, the results obtained using the intensity/energy (I/E)-based indices are shown as solid lines, while those based on the eigenvalue methods---particularly COMEDIE---are shown as dashed lines. 
%The eigenvalue-based results show only small deviations across all frequency bands. 
%A similar tendency to that seen in Figs.~\ref{fig:Psi_IE_AVE} and \ref{fig:angle_error} is observed for the {\bs Afmt} system, especially in the I/E-based indices, indicating that the frequency range with small estimation error is relatively narrow. 
%In contrast, even for {\bs Afmt}, the indices derived from the eigenvalues of the velocity covariance matrix exhibit comparatively small errors.

\begin{figure}[t] 
  \centering
  \includegraphics[width=0.9\linewidth,pagebox=artbox]{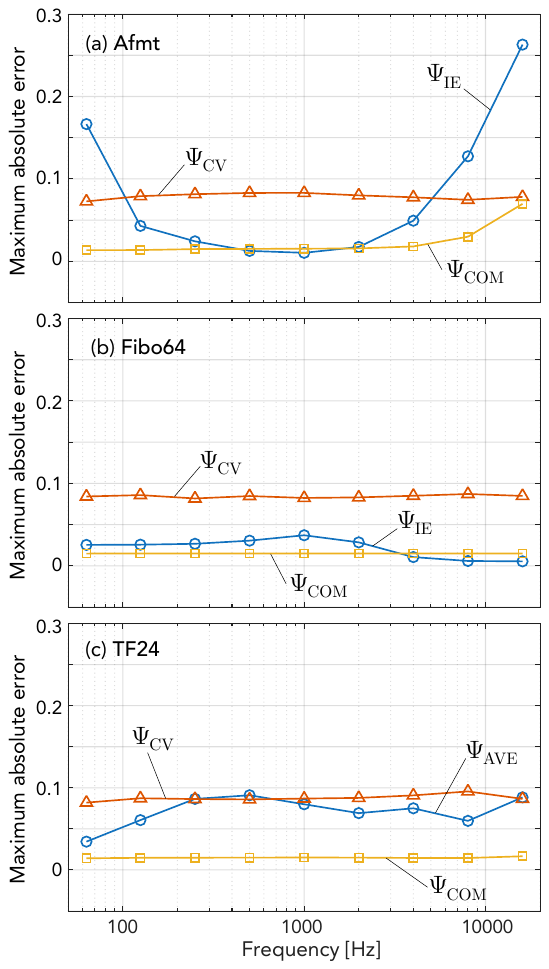}
 \caption{Case 2 (numerical simulation): Maximum absolute deviation between the beam ratio $1-\eta$ and the estimated diffuseness.
Results are shown for (a) {\bs Afmt}, (b) {\bs Fibo64}, and (c) {\bs TF24} configurations.
For each case, the intensity-based diffuseness measures $\Psi_{\mathrm{IE}}$ (or $\Psi_{\mathrm{AVE}}$ for {\bs TF24}), the coefficient-of-variation-based measure $\Psi_{\mathrm{CV}}$, and the eigenvalue-distribution-based measure $\Psi_{\mathrm{COM}}$ are compared as a function of frequency. }
  \label{fig:ratio_error}
\end{figure}

Figure \ref{fig:ratio_error} presents the results for (a) {\bs Afmt}, (b) {\bs Fibo64}, and (c) {\bs TF24}.
For each configuration, the intensity-based diffuseness measures $\Psi_{\mathrm{IE}}$ (or $\Psi_{\mathrm{AVE}}$ in the {\bs TF24} case), 
the coefficient-of-variation-based measure $\Psi_{\mathrm{CV}}$, and the eigenvalue-based measure $\Psi_{\mathrm{COM}}$ are shown.

For the {\bs Afmt} configuration, the intensity-based approach exhibits large deviations at both low and high frequencies.
In contrast, the error associated with the eigenvalue-based measure $\Psi_{\mathrm{COM}}$ remains small over the entire frequency range.
The $\Psi_{\mathrm{CV}}$ measure also shows stable behavior, with deviations below 0.1 across all frequency bands.

The same overall tendency is observed for the {\bs Fibo64} and {\bs TF24} configurations.
In all cases, $\Psi_{\mathrm{COM}}$ yields the smallest error, $\Psi_{\mathrm{CV}}$ remains consistently below 0.1, 
and the intensity-based measures $\Psi_{\mathrm{IE}}$ or $\Psi_{\mathrm{AVE}}$ exhibit intermediate performance.

\subsection*{Case~3: Reactive (interference) field}

\begin{figure}[t] 
  \centering
  \includegraphics[width=0.9\linewidth,pagebox=artbox]{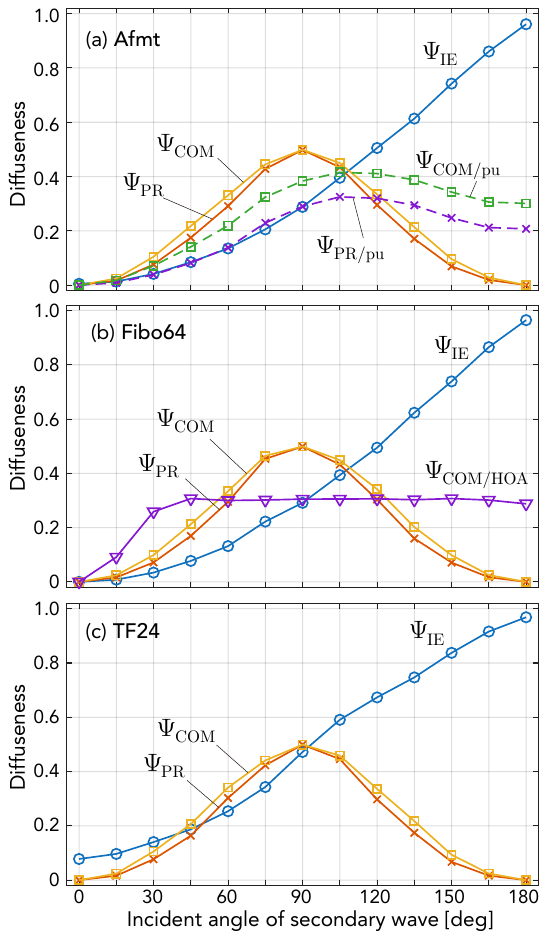}
 \caption{Case 3 (numerical simulation): Diffuseness for the interference field formed by two plane waves at various incident angles at \SI{1}{kHz}. 
(a)~{\bs Afmt}, (b)~{\bs Fibo64}, and (c)~{\bs TF24}. 
The results for $\Psi_{\mathrm{IE}}$, $\Psi_{\mathrm{PR}}$, and $\Psi_{\mathrm{COM}}$ are shown. 
For reference, in (a), the results of $\Psi_{\mathrm{PR}}$ and $\Psi_{\mathrm{COM}}$ obtained using all four FOA components ($p$ and $\bm{u}$) are also presented. 
In (b), the results of $\Psi_{\mathrm{COM}}$ calculated from the HOA coefficients (up to 4th) are additionally shown.
}
  \label{fig:two_waves}
\end{figure}

Figure~\ref{fig:two_waves} shows the diffuseness results for Case 3, in which two plane waves interfere at various incidence angles. The results are for 1kHz. 
Panels (a), (b), and (c) correspond to the results for {\bs Afmt}, {\bs Fibo64}, and {\bs TF24}, respectively. 
For all microphone arrays, both the intensity/energy-based index $\Psi_{\mathrm{IE}}$ and the eigenvalue-based indices $\Psi_{\mathrm{PR}}$ and $\Psi_{\mathrm{COM}}$ are presented. 
For reference, in (a) ({\bs Afmt}), $\Psi_{\mathrm{PR/pu}}$ and $\Psi_{\mathrm{COM/pu}}$, which were computed from the eigenvalues of the $4\times4$ covariance matrix including both pressure and velocity components after whitening, are also shown. 
In (b) ({\bs Fibo64}), $\Psi_{\mathrm{COM/HOA}}$, obtained from the eigenvalues of the whitened covariance matrix of the HOA coefficients, is additionally presented.

A notable feature is that the $\Psi_{\mathrm{IE}}$ index approaches unity as the incidence angle between the two plane waves approaches 180°, that is, when two plane waves of the same frequency arrive from opposite directions and interfere. 
This high diffuseness value arises because the acoustic intensity vector becomes nearly zero due to destructive interference. 
However, from a physical standpoint, such a condition cannot be regarded as a truly diffuse sound field, since the waves propagate only along a single axis (the $z$-direction in this case).

In contrast, for the eigenvalue-based indices calculated using only the velocity components, the diffuseness reaches its maximum value of approximately 0.5 at an incidence angle of 90°, and then gradually decreases as the angle increases, converging to zero at 180°. 
This tendency is almost identical for both $\Psi_{\mathrm{PR}}$ and $\Psi_{\mathrm{COM}}$. 
(Although not shown here for simplicity, if other eigenvalue-based measures such as the maximum-eigenvalue ratio or entropy-based metrics are used, the resulting diffuseness values fluctuate around these trends.) 
In panel (a), when the pressure component is also included in the covariance matrix, the peak position shifts, and the diffuseness converges to a value around 0.2 to 0.3 at 180°.

For the {\bs Fibo64} array shown in panel (b), the diffuseness index based on the eigenvalues of the HOA coefficients, presented for reference, converges to an approximately constant value of 0.3 for incidence angles above about 45°. 
When the interference of two plane waves at 180° is interpreted as representing a single propagation direction, it is reasonable that the diffuseness converges to zero. 
This behavior can be interpreted as supporting the validity of estimating diffuseness from the eigenvalue analysis of the velocity covariance matrix alone.

\begin{figure*}[t] 
  \centering
  \includegraphics[width=0.9\textwidth,pagebox=artbox]{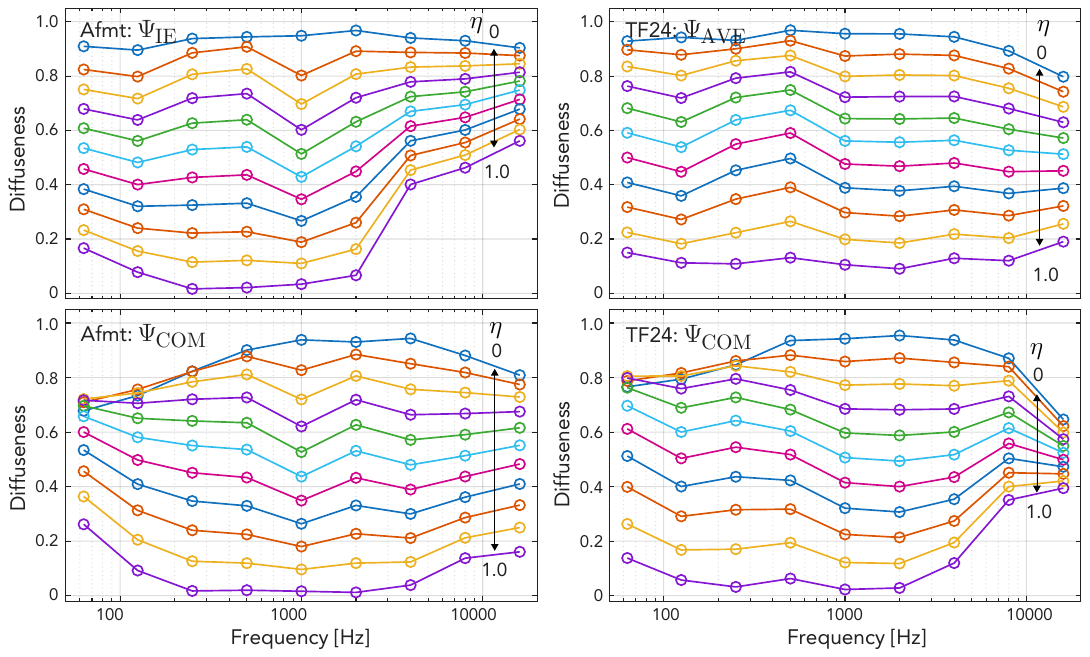}
 \caption{Diffuseness calculated from numerical simulations based on measured impulse responses,
obtained by mixing impulse responses measured in an anechoic room and a reverberation room
with various energy ratios.
The energy ratio of the anechoic component was varied from 0 to 1 in increments of 0.1.
The left panels show the results for the {\bs Afmt} array, while the right panels show those
for the {\bs TF24} array.
The upper panels correspond to the intensity-energy (I/E) method, and the lower panels show
the eigenvalue-based results using the COMEDIE index.}
  \label{fig:exp_ratio}
\end{figure*}

\section{Experimental verification}

\subsection{Concept}
Finally, the trends obtained from the simulations were experimentally verified using measured data. 
Two sound fields were considered: an anechoic chamber and a reverberation chamber. 
As expected, diffuseness is nearly zero in the anechoic chamber and close to unity in the reverberation chamber. 
However, unlike the numerical simulations, neither environment can be regarded as perfectly ideal, and it is difficult to ensure that the measured results precisely match these theoretical values. 
Therefore, to enable a meaningful verification, we adopted an approach analogous to Case 2 of the benchmark simulations: 
the impulse responses measured in the anechoic and reverberation rooms were used to represent the ``beam'' and ``diffuse'' components, respectively. 
By mixing these two responses at arbitrary energy ratios, we examined the linearity between the prescribed ratio and the resulting diffuseness.

Specifically, for both sound fields, impulse responses were measured using the {\bs Afmt} and {\bs TF24} microphone arrays. 
The energy within each frequency band was computed and combined according to the specified ratio, and the diffuseness was then calculated from the resulting signals. 
It should be noted that an Eigenmike was not available, and thus the case corresponding to the {\bs Fibo64} array was not experimentally validated.

\subsection{Conditions}

For the {\bs Afmt} array, an AMBEO VR Mic was used, and for the {\bs TF24} array, DPA 2012 microphones were employed with a spacing of 20 mm between each microphone pair. 
%
%As a preliminary measurement conducted in an anechoic room to determine the acoustic reference point of the shotgun microphone, impulse responses were measured twice at the same position: once with the microphone oriented toward the sound source, and once with it oriented perpendicularly.
%The time difference between the peaks of these two responses was used to estimate the reference point, which was found to be approximately 61 mm from the bottom of the microphone (amplifier section). Using this reference, 
The array was mounted on a rotatable jig (a camera tripod head) that allowed precise rotations in 45° increments around a point located 10 mm above the estimated center.

In the anechoic chamber, measurements were conducted at a distance of 2.5 m from the sound source. In the reverberation chamber (147 m$^3$, RT $\sim$ 4.7 s), the distance to the source, which was directed toward a wall, was approximately 5 m.

\subsection{Results}

The results are shown in Fig.~\ref{fig:exp_ratio}. 
The left panels correspond to the {\bs Afmt} array, and the right panels to the {\bs TF24} array. 
The upper panels show the diffuseness obtained using the intensity-energy (I/E) method, while the lower panels present the diffuseness $\Psi_{\mathrm{COM}}$ calculated from the eigenvalues of the covariance matrix constructed only from the velocity components of the FOA signals. 

%It is evident that the results for the {\bs Afmt} array exhibit a larger deviation from $1-\eta$ in both low and high frequency ranges, and the overall trend appears distorted. 
%In contrast, the {\bs TF24} array demonstrates a more linear relationship. 
In this simulation, the $\Psi_{\mathrm{AVE}}$ index obtained with the {\bs TF24} array exhibits an overall stable and approximately linear variation with respect to the mixing ratio.
In contrast, the I/E-based results for the {\bs Afmt} array show relatively high diffuseness values above 4kHz, while the eigenvalue-based $\Psi_{\mathrm{COM}}$ for {\bs Afmt} displays a slightly biased behavior in the low-frequency range.
Furthermore, for the {\bs TF24} array, the $\Psi_{\mathrm{COM}}$ index also deviates from a linear trend at 8kHz and 16~kHz under the present simulation conditions.

The fact that the diffuseness does not reach exactly unity when $\eta=0$ (reverberation room only) and does not converge to zero when $\eta=1$ (anechoic room only) is attributed to both the non-ideal characteristics of the measured sound fields and the inevitable experimental errors, so the absolute values should be interpreted with appropriate caution.

\begin{figure}[t] 
  \centering
  \includegraphics[width=0.9\linewidth,pagebox=artbox]{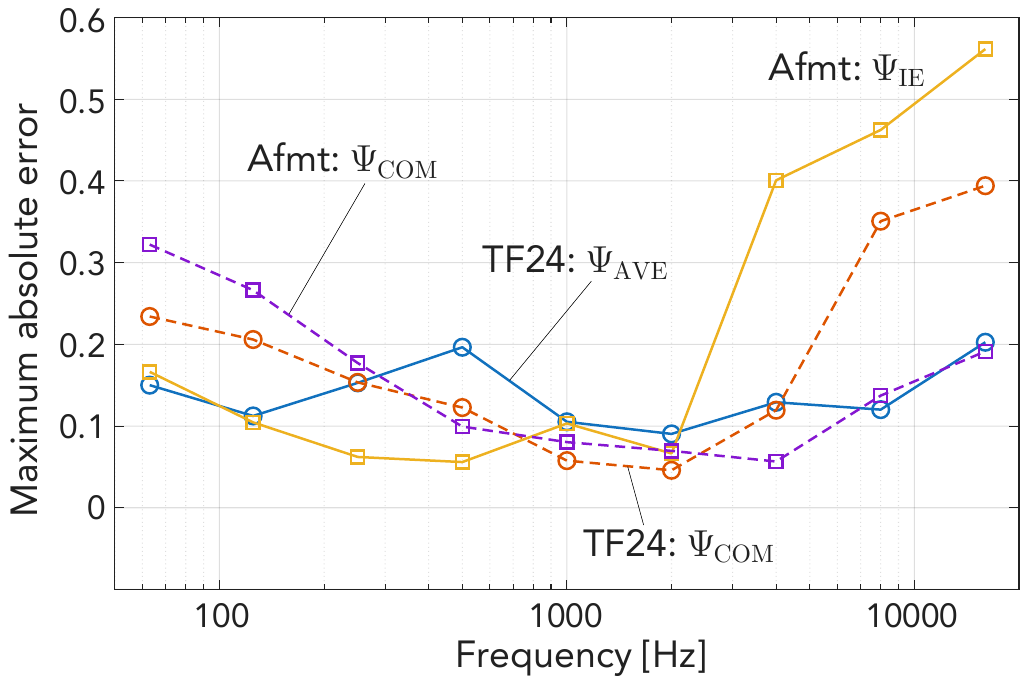}
 \caption{The maximum absolute difference between the calculated diffuseness and $1-\eta$ for each index, in which the diffuseness was calculated from numerical simulations based on measured impulse responses}
  \label{fig:exp_error}
\end{figure}

As a reference, Fig.~\ref{fig:exp_error} shows the maximum absolute difference between the calculated diffuseness and $1-\eta$ for each index. 
The overall trends are consistent with those in Fig.~\ref{fig:ratio_error}: 
the intensity-energy (I/E) indices generally yield slightly larger errors than the eigenvalue-based indices, and the {\bs Afmt} array exhibits larger deviations than the {\bs TF24} array.

\section{Conclusion}

This study presented a comparative investigation of diffuseness estimation methods based on intensity-energy relations and eigenvalue analysis, using several microphone arrays including the {\bs Afmt}, {\bs Fibo64}, and {\bs TF24} systems. 
Through a series of controlled simulations, the relationships between the diffuseness index and the beam-to-diffuse ratio were systematically examined under single-source, mixed, and interference conditions. 
The results demonstrated that the eigenvalue-based indices, particularly $\Psi_{\mathrm{COM}}$, exhibited higher robustness and a more linear response than the conventional intensity-energy (I/E) approach. 
The I/E method tended to overestimate diffuseness in interference fields where the active intensity approaches zero, which may lead to physically inconsistent interpretations.

Regarding the combination of microphone array configurations and evaluation approaches, the {\bs Fibo64} array would be the first choice if such an array is available. In cases where this is not realistic, arrays composed of microphones with known directivity patterns---such as the Tight frame configuration, which possesses theoretical well-defined structure---represent a reasonable alternative. Furthermore, the eigenvalue analysis based on the covariance matrix constructed only from the FOA velocity components ($u_x$, $u_y$, and $u_z$) was found to produce stable and reliable results overall.

Experimental verification using impulse responses measured in anechoic and reverberation rooms further confirmed the general trends predicted by the simulations. 
The diffuseness estimated from mixed impulse responses showed approximately linear dependence on the energy ratio between the direct and reverberant components, with smaller deviations observed for the {\bs TF24} array compared with the {\bs Afmt} array. 
These results suggest that the velocity-based covariance analysis provides a practical and reliable framework for characterizing the spatial diffuseness of sound fields, even without strict whitening or mode-power normalization.

While the {\bs TF24} formulation does not aim at a strict physical equivalence to pressure-velocity measurements, the present study demonstrates that, once its assumptions and limitations are explicitly stated, the proposed pseudo-intensity-based formulation provides a practically useful and internally consistent basis for comparative diffuseness evaluation.
In this sense, the {\bs TF24} framework should be interpreted not as a replacement for physically exact intensity measurements, but as an operational representation that enables stable and interpretable diffuseness estimation across different array geometries.

Future work will extend the present analysis beyond the weighted averaging of directional diffuseness, exploring how this parameter relates to perceptual impressions of spatial quality. Further investigations will aim to establish practical applications of diffuseness-based metrics in spatial reproduction and acoustic design, where such perceptually informed measures may provide useful guidance for sound field control and evaluation.

\section*{ACKNOWLEDGMENTS}
The author would like to thank Mr. T. Ohno, Mr. R. Ohnuki, and Mr. K. Hirota, a graduate of his laboratory, for their contributions during the early stages of this research and measurements. 
The author also thanks Dr. H. Kashiwazaki for his valuable suggestions and insightful comments regarding the Tight frame structure.
This research was supported by JSPS KAKENHI Grant No. JP24K03222.

%%%%%%%%%%%%%%%%%%%%%%%%%%%%%%%%%%%%%%%%%

\section*{Author Declarations}

\noindent\textbf{Conflict of Interest:}  
The author declares no conflicts of interest.

\medskip
\noindent\textbf{Data Availability:}  
The data that support the findings of this study are available from the corresponding author upon reasonable request.

\medskip
\noindent\textbf{Ethics Approval:}  
Not applicable.

%%%%%%%%%%%%%%%%%%%%%%%%%%%%%%%%%%%%%%%%%

\vspace{3mm}

\noindent {\small
a) mh acoustics Eigenmike em64 spherical microphone array. 
Available at: \url{https://eigenmike.com/eigenmike-64} (Accessed October 24, 2025). \\
b) Sennheiser AMBEO VR Mic, 3D audio microphone. 
Available at: \url{https://www.sennheiser.com/en-us/catalog/products/microphones/ambeo-vr-mic/} (Accessed October 24, 2025).}

\end{document}